\documentclass[pra,aps,12pt,a4paper]{revtex4-1}
\usepackage{epsfig,rotating}
\usepackage{graphicx}
\usepackage{epstopdf,color}
\usepackage{amsmath,bm}
\usepackage{color}
\usepackage[colorlinks=true,urlcolor=blue,linkcolor=blue,citecolor=blue]{hyperref}
\newcommand{\average}[1]{\left\langle #1 \right\rangle}
\newcommand{\half}{\frac{1}{2}}
\begin{document}

\title{High-order strong methods for stochastic differential equations with colored noises}

\author{Shuanglin Sun} 
\affiliation{School of Business, Ludong University, Shandong 264025, China}
\author{Yun-An Yan}\email{yunan@ldu.edu.cn}
\affiliation{School of Physics and Optoelectronic Engineering,
  Ludong University, Shandong 264025, China}

\begin{abstract}
The key difficulty to develop efficient high-order methods for integrating
stochastic differential equations lies in the calculations of the 
multiple stochastic integrals. This letter suggests a scheme to compute 
the stochastic integrals for the colored noises based on the white noise 
representation. The multiple stochastic integrals involving one 
and two stationary noises can be conveniently generated 
together with noises using the discrete Fourier transformation.
Based on the calculated stochastic integrals, we obtain simple fourth-order 
and third-order strong methods for equations with a single and multiple noises,
respectively. Numerical tests verify the accuracy of the suggested methods.

\end{abstract}

\maketitle

\section{Introduction}

We consider the initial value problems with the linear 
stochastic differential equations (SDEs)
\begin{eqnarray}
  \label{eq:sde}
  \frac{d \bm{X}(t)}{dt} = A \bm{X}(t) + \sum_{j=1}^K B_j \bm{X}(t) \xi_j(t)
\end{eqnarray}
where $\bm{X}(t)$ is an $M$-dimensional state vector, 
$A$ and $B_j$ are $M\times M$ matrices, 
$K$ is the number of noises, 
and $\xi_j(t)$ are stationary colored Gaussian noises with zero means
$\average{\xi_j(t)}=0$ and two-time correlation functions 
$\average{\xi_j(t)\xi_k(s)} = \alpha_{jk}(t-s)$. 
Here we focus on the noises with continuous correlation functions $\alpha_{jk}(t)$.

Linear SDEs arise in simulating the dynamics of social or physical 
systems perturbed by noises. For open physical systems, the 
dissipative dynamics is often described by system-plus-environment model and the 
effect of the environment on the evolution of the system can be characterized by
noises. With such a paradigm, the stochastic approach becomes a powerful
tool to study quantum dissipative dynamics as well as 
quantum measurements~\cite{gisin92_5677,carmichael93}.
Various stochastic methods are available in the literature, such as
the stochastic Liouville equation~\cite{stockburger98_2657,shao04_5053,yan15_022121} and
the stochastic Schr\"{o}dinger's 
equation~\cite{cao96_4189,strunz97_25,strunz99_1801,jing15_022109,ke16_024101}.
As most stochastic simulations focused on dynamics of bosonic systems,
the stochastic description was recently extended to fermionic quantum 
dissipation~\cite{han19_050601}. For social systems, linear SDEs are used 
to simulate the stock price in the Black-Scholes model and 
become the basis for mathematical finance~\cite{Lamberton96}.

Due to the wide applications and the importance of SDEs, it has attracted extensive 
attentions to  develop efficient numerical integrators for stochastic equations.
However, high-order integrators have to deal with the multiple stochastic integrals 
and become more sophisticated than its deterministic counterpart.
To reduce the complexity in stochastic integration, 
the average-converged weak methods were suggested as useful alternatives to
the pathwise strong methods~\cite{kloeden92_283,kloeden89_155,kloeden02_187,abdulle08_997}. 
With the Brownian path, one can also develop variable step-size 
approaches~\cite{gaines97_1455,mauthner98_93,burrage04_317,ilie15_791}.
In order to obtain simpler integrators,
the idea of Runge-Kutta methods originally designed 
for ordinary differential equations were also applied to the stochastic 
cases~\cite{rossler10_922,burrage04_373,wang08_324,bastani07_631,xiao16_259}.

For linear SDEs, the linearity allows us to develop more 
efficient and yet simpler integrators. For example, while strong order 1 Milstein 
method~\cite{milstein74_583} is widely used and high strong order 
integrators become sophisticated for general SDEs~\cite{burrage04_373}, 
high strong order schemes for linear equations are much simpler based on the Magnus 
formula~\cite{burrage99_34,lord08_2892}.
When the noise is of the Ornstein-Uhlenbeck type, we can even derive efficient, deterministic  
hierarchical approaches to solve the stochastic average 
$\average{\bm{X}(t)}$~\cite{shapiro78_563,yan04_216}.

In simulating physical systems, one often needs to use colored noise to represent
the effect of the environment~\cite{stockburger98_2657,
shao04_5053,yan15_022121,strunz97_25,strunz99_1801,jing15_022109,ke16_024101}. 
In this case, one can of course use the integrators designed for the white noises 
to simulate the dynamics. But more efficient schemes could be developed by 
utilizing the noise correlation. For example, Milshtein and Tret'yakov suggested 
a 5/2 strong order method for colored noises~\cite{milshtein94_691}. 
Honeycutt even put forward a fourth-order strong method
for Ornstein-Uhlenbeck noises~\cite{honeycutt92_604}.

In this letter, we take advantage of the correlation 
for the colored noises and suggest an efficient way to 
calculate the multiple stochastic integrals with the 
discrete Fourier transformation. The current approach 
results in a simple third-order strong integrator for SDEs 
with multiple colored noises and a fourth-order method 
for SDEs with a single noise. 

\section{Calculations of multiple stochastic integrals for colored noises}

For simplicity, we extend lower bound of the summation in Eq.~(\ref{eq:sde}) to zero and 
rewrite the equation as 
$\dot{\bm{X}} = \sum_{j=0}^K B_j \bm{X}(t) \xi_j(t)$ 
so that $B_0\equiv A$ and $\xi_0(t)=1$.
In numerical simulations, we normally discretize the time interval $[0,T_f]$
with a uniform grid $\{t_0,t_1,\cdots,t_N\}$ where $N$ is the 
number of time steps, $\Delta t$ is the step size, and $t_j=j\Delta t$.
For linear SDEs, one can employ the Magnus expansion
to find an approximation for $\bm{X}(t+\Delta t)$ 
based on $\bm{X}(t)$~\cite{magnus54_649,burrage99_34},
\begin{equation}
  \label{eq:magnus}
  \bm{X}(t+\Delta t) = e^{\Omega(t, \Delta t)}\bm{X}(t). 
\end{equation}
Up to the third order, $\Omega(t, \Delta t)$ reads
\begin{eqnarray}
  \hspace{-0.5cm}\Omega(t, \Delta t) = \sum_j B_j J_j(t) 
  +\half\sum_{j<k} [B_j, B_k] K_{kj}(t)
  + \frac{1}{6}\sum_{j,k<l}[B_j, [B_k, B_l]] K_{lkj}(t),
\end{eqnarray}
where $J$s represents the multiple stochastic integral,
\begin{eqnarray}
  &&\hspace*{-1.1cm} J_j(t)    = \int^{t+\Delta t}_t dt_1 \xi_j(t_1),\\
  &&\hspace*{-1.1cm} J_{jk}(t) = \int^{t+\Delta t}_t dt_1 \xi_k(t_1)\int^{t_1}_t dt_2 \xi_j(t_2),
  \\ &&\hspace*{-1.1cm}
  J_{jkl}(t) = \int^{t+\Delta t}_t dt_1 \xi_l(t_1)\int^{t_1}_t dt_2 \xi_k(t_2)
    \int^{t_2}_t dt_3 \xi_j(t_3),
\end{eqnarray}
and
\begin{eqnarray}
  \label{eq:me2}
  K_{kj}(t) &=& J_{kj}(t)-J_{jk}(t),\\
  \label{eq:me3}
  K_{lkj}(t) &=& J_{lkj}(t)-J_{jlk}(t)-J_{klj}(t)+J_{jkl}(t).
\end{eqnarray}
Note that for white noises, one should be cautious about the calculations of 
multiple stochastic integrals because the It\^o integration and the Stratonovich integration will
yield different results. But here these two methods will 
produce the same results since $\xi_j(t)$ are continuous functions of time.

The stationary colored noises can be represented with white noises,
that is,
\begin{eqnarray}
  \label{eq:rep}
  \xi_{j}(t) = \Re \int dW(\omega) e^{-i\omega t}\Gamma_j(\omega), 
\end{eqnarray}
where $W(\omega)$ is a Wiener process in the frequency domain. 
In this letter, the integration range is always $[-\infty,\infty]$ 
with respect to the Wiener process $W(\omega)$ and omitted in the expression. 
One may verify that $\xi_j(t)$ satisfies the desired correlations
if $\Gamma_j(\omega) \Gamma^\ast_k(\omega)$ is 
the inverse Fourier transformation
of $\alpha_{jk}(t)$. In practice, the circulant embedding 
method of the correlations may be used to solve the discretized approximation 
for $\Gamma_j(\omega)$~\cite{chan99_265,yan16_110309}.

\begin{figure}[tb!]
  \centering
   \includegraphics[width=0.85\textwidth]{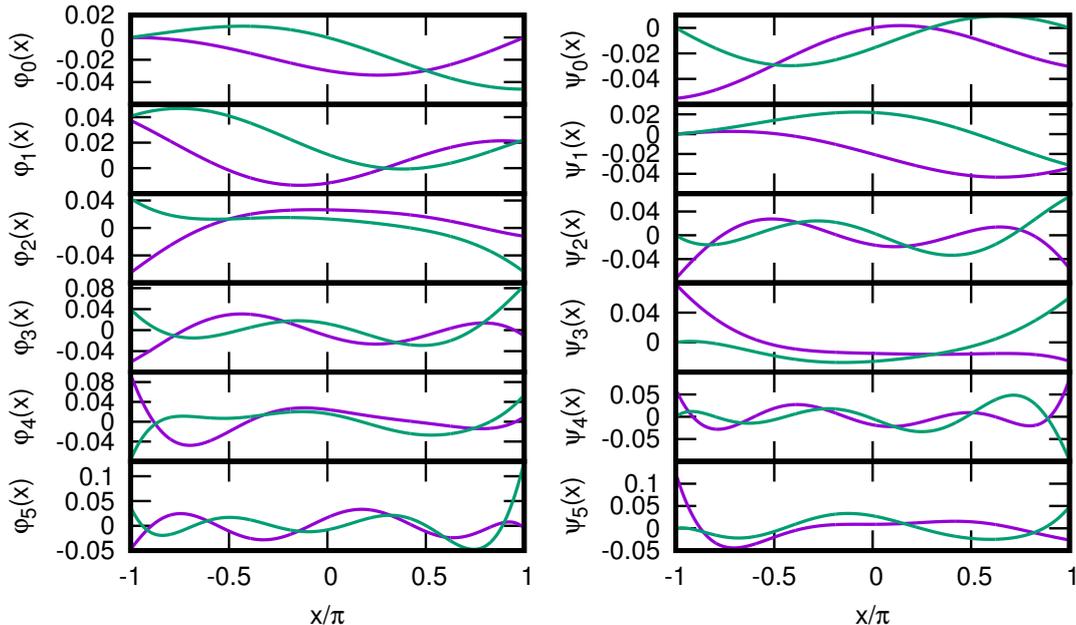}
\caption{\label{fig:svd} The first six eigenfunctions $\varphi_j(x)$ and $\psi_j(x)$ 
  from the singular value decomposition of $\mathcal{H}(x, y)$ 
  after discretized with a 1024-point grid. The real and imaginary parts 
of the functions are displayed in purple and green, respectively.
The eigenvalues $\epsilon_0$-$\epsilon_5$ are
$2.475\times 10^{2}$,
$2.476\times 10^{2}$,
$2.680\times 10^{-1}$,
$2.680\times 10^{-1}$,
$1.746\times 10^{-5}$, and
$1.746\times 10^{-5}$, respectively. 
The next eigenvalue $\epsilon_6=2.435\times10^{-10}$ is five order smaller than 
$\epsilon_5$.
}
\end{figure}

With the noise representation, one can directly calculate the 
multiple integrals involved in Eq.~(\ref{eq:magnus}). 
In a third-order simulation, $J_j$ and $J_{jk}$ are 
all the needed multiple stochastic integrals, and can be 
calculated with the white noise representation of $\xi_j(t)$ for $j\neq 0$,
\begin{eqnarray}
  \label{eq:Js}
  &&\hspace{-1cm}J_j(t) = \Im \int dW(\omega) e^{-i\omega t}\frac{\Gamma_j(\omega)}{\omega}
    \big[1-e^{-i \omega\Delta t}\big],\\
  \label{eq:Jj0}
  &&\hspace{-1cm}K_{j0}(t) =\Re\int dW(\omega) e^{-i\omega t} \frac{\Gamma_j(\omega)}{\omega^{2}}\big[2-i \omega\Delta t
  -e^{-i \omega\Delta t}(2+i \omega\Delta t)\big],\\
  \label{eq:Jjk}
    &&\hspace{-1cm}K_{jk}(t) =\Re \int dW(\omega_1)\int dW(\omega_2) \frac{\Delta t^2}{2}\Gamma_j(\omega_1)
    \nonumber \\ && \quad \times
  e^{-i\omega_1 t} \Big[\mathcal{H}(\omega_1 \Delta t, \omega_2 \Delta t) \Gamma_k(\omega_2) e^{-i\omega_2 t}
  +\mathcal{H}(\omega_1 \Delta t, -\omega_2 \Delta t) \Gamma_k(-\omega_2) e^{i\omega_2 t}\Big ],
\end{eqnarray}
where $\mathcal{H}(\omega_1,\omega_2)$ is the kernel 
\begin{eqnarray}
  &&\hspace{-1cm} \mathcal{H}(x, y) = \frac{e^{-i (x+y)}}{x y (x+y)}
    \Big[ x (1+e^{i x})(1-e^{i y})
     -y (1-e^{i x})(1+e^{i y})
  \Big].
\end{eqnarray}
In numerical simulations, we usually employ the discrete Fourier 
transformation to calculate the integrals. Then the computational
effort will be $O(N^2\log(N))$ for a straightforward implementation of Eq.~(\ref{eq:Jjk}).
To reduce the computational effort in noise generation, 
we approximate $\mathcal{H}(x,y)$ with a degenerate kernel,
\begin{eqnarray}
  \mathcal{H}(x,y) \approx \sum_{j=0}^{N_f-1} \epsilon_j \psi_j(x) \varphi_j(y),
\end{eqnarray}
where $N_f$ is the number of terms for the approximation, 
$\epsilon_j$, $\psi_j(x)$ and $\psi_j(x)$ are the eigenvalues, 
the left and right eigenfunctions of the kernel, respectively.
Note that here the vector elements are counted from 0.
The degenerate kernel can be obtained with a singular value decomposition 
after discretizing $\mathcal{H}(x,y)$.
As Fig.~(\ref{fig:svd}) shows, a six-term kernel can well approximate $\mathcal{H}(x,y)$ 
and the maximum absolute error will be as low as the order of $10^{-12}$. 
With such a degenerate kernel approximation, the computational effort
for producing $J_{jk}-J_{kj}$ is only six times of that for the noise.

Note that $K_{lkj}(t)$ is of the fourth order with respect to $\Delta t^4$.
We thus can construct a third-order strong method for SDEs with colored noises 
with a second-order Magnus expansion. 
For SDEs with a single noise, the $O(\Delta t^4)$ multiple stochastic integrals 
are $K_{100}(t)$ and $K_{101}(t)$,
which can also be calculated with Eq.~(\ref{eq:rep}). The result 
for $K_{100}(t)$ reads
\begin{eqnarray}
  \label{eq:J100}
&&\hspace{-1cm}K_{100}(t) =\Re\int dW(\omega) \frac{e^{-i\omega t}}{2 \omega^3}
  \big[
    6 \omega \Delta t (1+e^{-i \omega \Delta t}) 
    + i (12-\omega^2 \Delta t^2) (1-e^{-i \omega \Delta t})
  \big]\Gamma_j(\omega).
\end{eqnarray}
The expression for $K_{101}(t)$ assume the same form as Eq.~(\ref{eq:Jjk}) 
for $K_{jk}(t)$, but with the kernel 
$\mathcal{H}(x,y)$ replaced by $\tilde{\mathcal{H}}(x,y) \Delta t$. 
The kernel $\tilde{\mathcal{H}}(x,y)$ reads
\begin{eqnarray}
  \label{eq:J101}
  \tilde{\mathcal{H}}(x,y) &=& 
  \frac{e^{-i (x+y)}}{x^2 y^2 (x+y)^2}
  \Big\{ 
      \big[e^{iy} (i + y) - i 
      + e^{ix} (y-i + i e^{iy})\big] x^3
      - 3 i (1 - e^{ix}) (1 + e^{iy}) x y^2
        \nonumber \\ &&
      + \big[1 + 2 e^{iy} + e^{ix} (2 + e^{iy})\big] x^2 y^2
      + \big[(1 + e^{iy}) (-2 i + x + e^{ix} (2 i + x))\big] y^3
    \Big\}.
\end{eqnarray}
Again, the kernel $\tilde{\mathcal{H}}(x,y)$ can be approximated 
with a degenerate one to reduce the computational efforts.

Upon the calculated stochastic integrals, Eq.~(\ref{eq:magnus}) is 
transformed into an ordinary differential equation 
\begin{eqnarray}
  \label{eq:ode}
  \frac{d\bm{X}(s)}{d s} = \mathcal{L}(t)\bm{X}(s)
\end{eqnarray}
for $t\leq s \leq t+\Delta t$. 
Note that the matrix $\mathcal{L}(t)=\Omega(t,\Delta t)/\Delta t$
is constant in the time interval $[t, t+\Delta t]$
so that Eq.~(\ref{eq:ode}) can be integrated with the low-storage Runge-Kutta methods 
specifically designed for autonomous linear equations~\cite{yan17_277}.
We thus obtain a third-order strong method for multiple-noise SDEs 
and a fourth-order strong method for single-noise SDEs.

\section{Numerical tests and discussion}

\begin{figure}[tb!]
  \centering
  \includegraphics[width=0.85\textwidth]{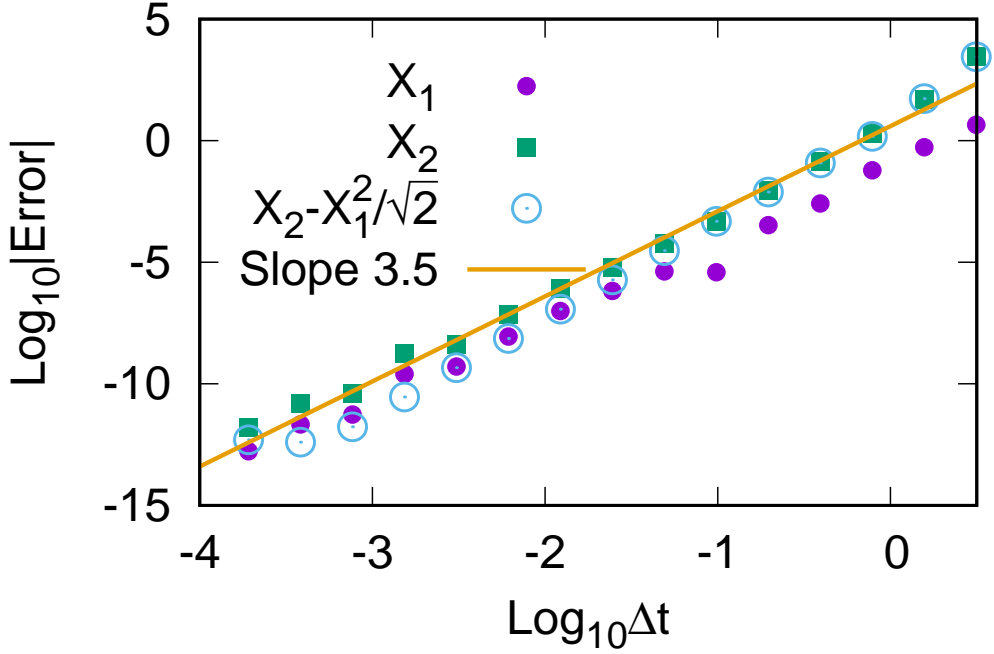}
  \caption{\label{fig:qho} Errors for the stochastically-driven 
    quantum harmonic oscillator with the third-order method.
  Bullets: the absolute errors of the amplitude in the first excited state;
  Squares: the absolute errors of the amplitude in the second excited state;
  Circles with dot: the absolute errors in the identity $X_2(t) - X_1^2(t)/\sqrt{2}$.
  The line with slope 3.5 is shown for guide of eye.
}
\end{figure}
\begin{figure}[tb!]
  \centering
  \includegraphics[width=0.45\textwidth]{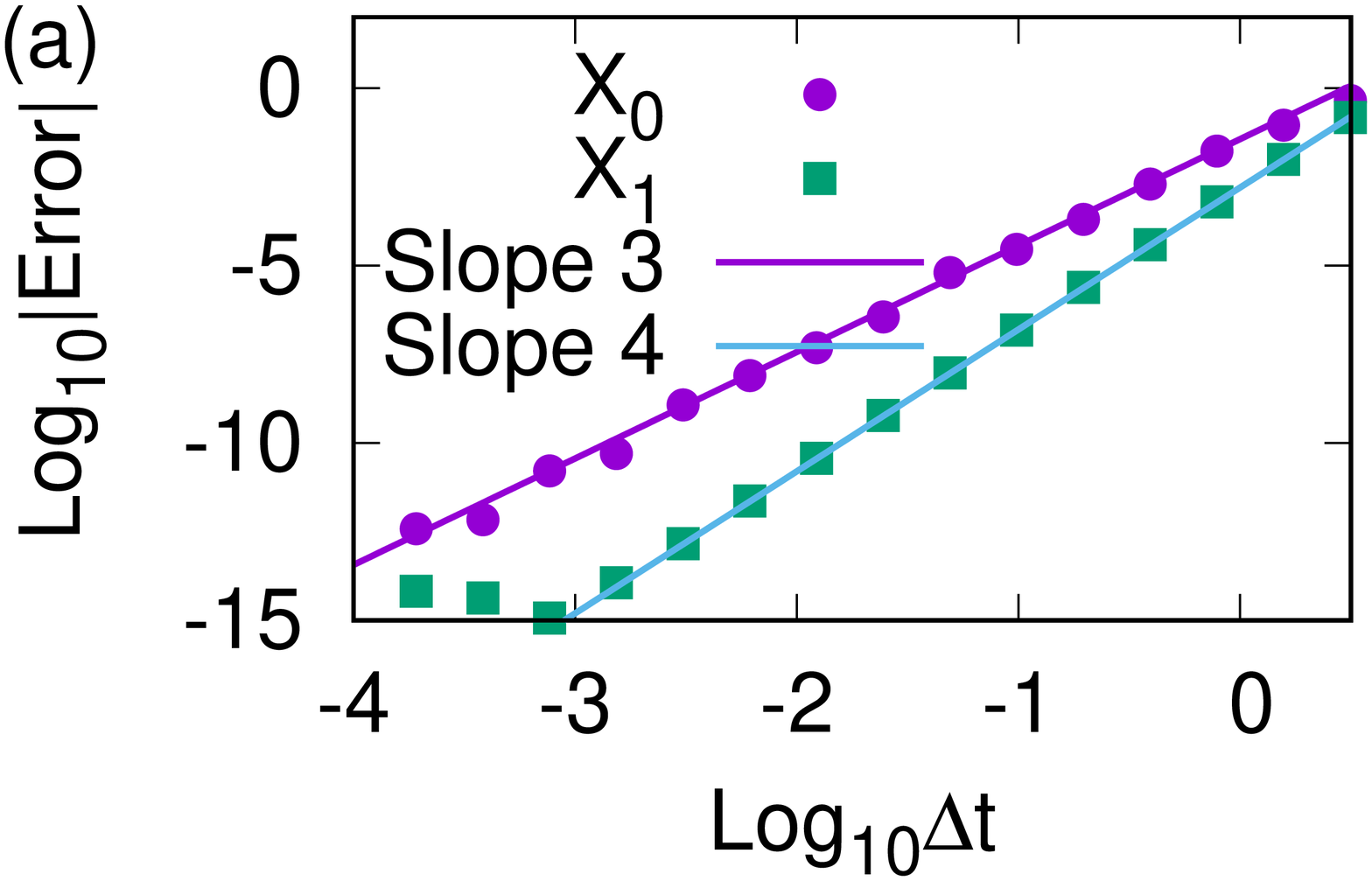}
  \includegraphics[width=0.45\textwidth]{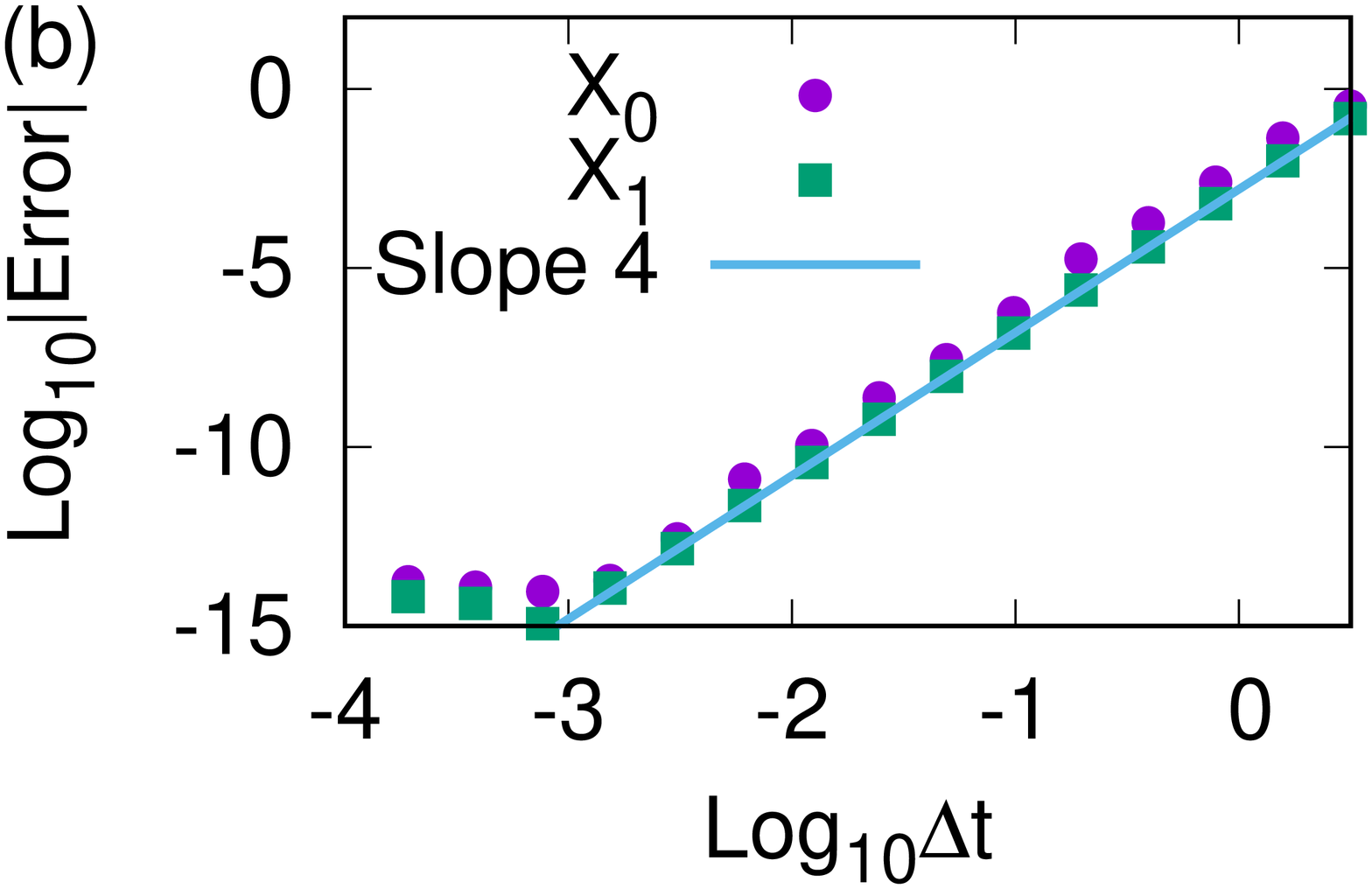}
  \caption{\label{fig:tls} Errors for the two-state system driven by 
    a single noise
    simulated with the third-order (a) and the fourth-order (b) methods.
  Bullets: the absolute errors for $X_0$;
  Squares: the absolute errors for $X_1$.
  Lines with slope 3 and 4 are also shown for guide of eye.
}
\end{figure}
\begin{figure}[tb!]
  \centering
  \includegraphics[width=0.85\textwidth]{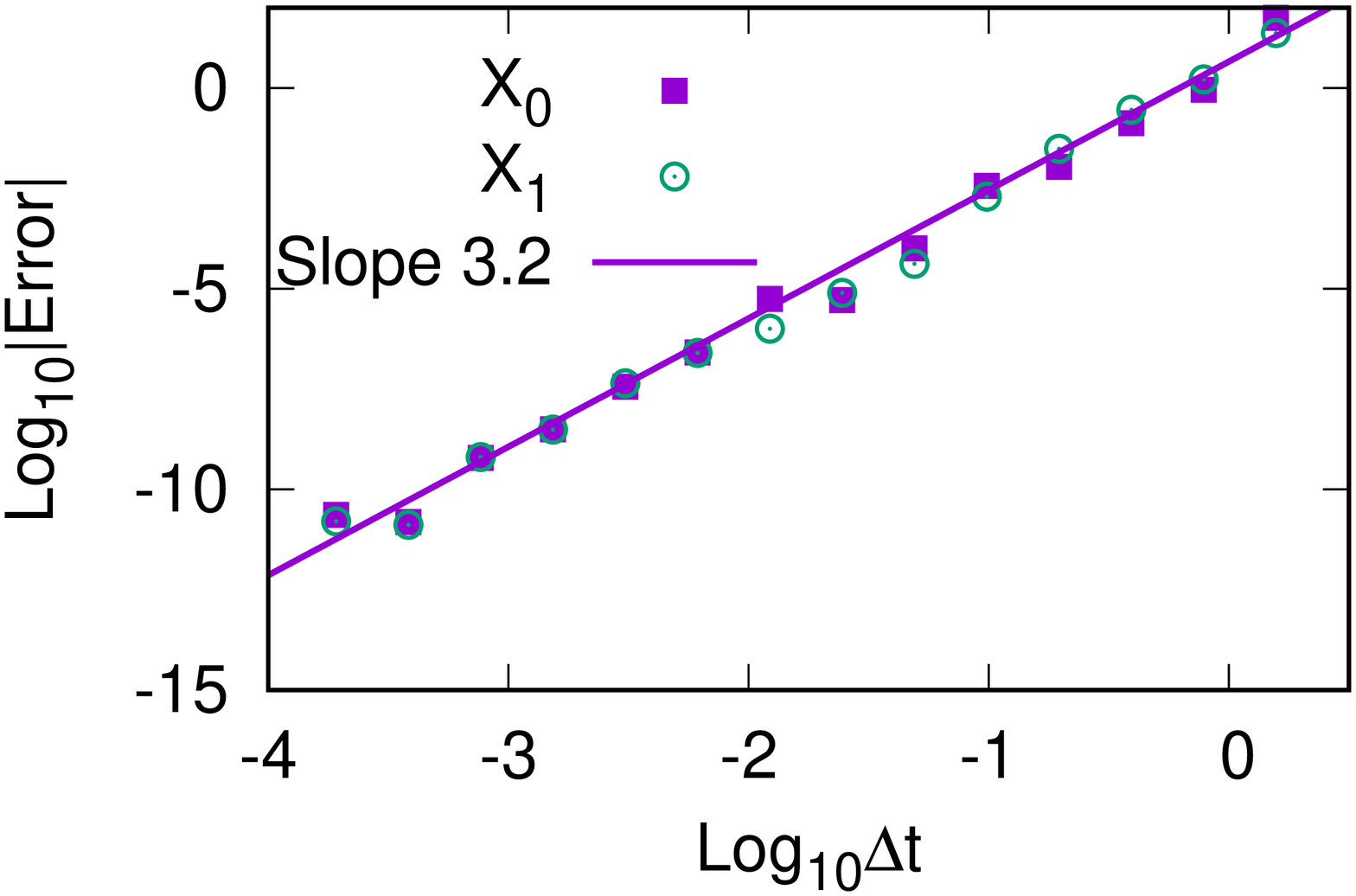}
  \caption{\label{fig:tls2n} Errors for the two-state system 
    perturbed by two noises.  Squares: the absolute errors for $X_0$;
  Circles with dot: the absolute errors for $X_1$.
  The line with slope 3.2 is shown for guide of eye.
}
\end{figure}

We will discuss the performance of the stochastic integrators 
with two uncorrelated, colored noises $\xi_1(t)$ and $\xi_2(t)$ satisfying the correlation
$\alpha_{11}(t) = 2 e^{-\pi t/2}$, $\alpha_{22}(t) = 5 e^{-\pi t/\sqrt{2}}$
and $\alpha_{12}(t)=0$.

The first numerical example is the stochastically-driven quantum harmonic oscillator  
with $A = -i \hat{a}^\dagger\hat{a}$ and $B_1 = -i\hat{a}^\dagger$ in Eq.~(\ref{eq:sde}). 
Here $\hat{a}^\dagger(\hat{a})$ is the creation (annihilation) operator
of the oscillator. The dynamics is propagated with the third-order method 
for the lowest 20 levels with the initial condition $X_0(0)=1$ and $X_j(0)=0\,(j\neq0)$.
This model is analytically solvable and the solutions are
$X_n(t) = e^{-i n (t + \frac{\pi}{2})} L^n_1(t,0)/\sqrt{n!},$
where $L_1(t,s) = \int^t_s d\tau\, \exp(i \tau)\xi_1(\tau)$. 
The function $L_1(t+\Delta t, t)$ may be generated together with the 
noise $\xi_1(t)$ and the multiple stochastic integrals, that is,
\begin{eqnarray}
  \label{eq:L1}
  L_1(t+\Delta t, t) &=& 
  \Re\int dW(\omega) e^{-i\omega t} \frac{\Gamma_1(\omega)}{\omega^2-1}
 \Big\{-i \omega 
 + \big[i \omega \cos(\Delta t) -\sin(\Delta t)\big] 
 e^{-i w\Delta t}\Big\} 
 \nonumber \\ && \quad \hspace{-0.5cm}
 -i \Re \int dW(\omega) e^{-i\omega t}  \frac{\Gamma_1(\omega)}{\omega^2-1}
 \Big\{\big[\cos(\Delta t) 
 + i \omega \sin(\Delta t) - 1\big] e^{-i \omega \Delta t}\Big\}.
\end{eqnarray}
We would like to stress that the same white noise $W(\omega)$ 
should be used to calculate the stochastic integrals 
in Eqs.~(\ref{eq:Js})-(\ref{eq:Jjk}) and (\ref{eq:L1}). Note that 
with the spirit of Eq.~(\ref{eq:L1}) we can apply the 
current method to propagate time-dependent SDEs. 

The stochastic equation is integrated to the final time $T_f = 2 \pi$ 
with $2^n$ ($1\leq n\leq 18)$ steps using the same stochastic trajectory. 
This task can be done in a recursive manner. First, a $2^{18}$-step propagation 
is performed with the stochastic integrals $J_1$, $K_{10}$ 
and $L_1$ generated from the discrete Fourier transformation.  
Then the step size is doubled and stochastic integrals with $2^{n-1}$ steps
are calculated based on those with $2^n$ steps.
The matrix exponential function in Eq.~(\ref{eq:magnus}) is calculated with 
the fourth-order low-storage Runge-Kutta method~\cite{yan17_277}.

We employ two methods to analyze the errors for $X_1(T_f)$ and $X_2(T_f)$. 
First, the results with the smallest step size are used as the reference. 
Second, the identity $X_2(t) = X^2_1(t)/\sqrt{2}$ for the solutions is utilized 
to verify the results. The errors are depicted in Fig.~\ref{fig:qho}, which clearly 
shows the errors are of $O(\Delta t^{3.5})$. 

The second example is the simplest non-commutative two-state system with
$A=-i\sigma_z/2$, $B_1=(\sigma_x+i\sigma_y)/2$ and the initial condition
$\bm{X}(0)= (0, 1)^T$.
Here $\sigma_x$, $\sigma_y$ and $\sigma_z$ are Pauli matrices.
This model is analytically solvable and the solutions are
$X_0(t) = e^{-i t/2} L_1(t,0)$ and $X_1(t) = e^{i t/2}$. 
Fig.~(\ref{fig:tls}) illustrates the errors for the two-state model with 
different step sizes. Because the evolution of $X_1(t)$ in this model is 
deterministic, its accuracy is controlled solely by the method to calculate 
the matrix exponential. The plots show that the errors are $O(\Delta t^4)$ as 
it should for a fourth-order method in calculating the matrix exponential.
Different from $X_1$, the propagation of $X_0(t)$ depends on the stochastic integrals,
whose errors change with the order of stochastic integrator.
Here we display the results with and without incorporating the 
triple stochastic integrals and the orders are four and three, respectively, 
which are the expected results.

The last example is a two-state system driven by two noises, whose
dynamics is described by $A=i\sigma_z$, $B_1=i\sigma_x$, and $B_2=i\sigma_y$.
For such a model, a fourth-order integrator is complicated and the
third-order strong method is used to propagate the system with 
the initial state $\bm{X}(0)= (1, 0)^T$. This model is not analytically solvable 
and the results with $2^{18}$ steps will serve as the reference.
The errors with different step sizes are presented in Fig.~\ref{fig:tls2n},
which shows that the errors approximately scale as $O(\Delta t^{3.2})$.

\section{Conclusions}
Magnus expansion is a useful tool to develop simple and efficient
strong methods for linear stochastic equations and then the 
key step is the calculations of multiple stochastic integrals.
Here we have suggested a method to generate the 
multiple stochastic integrals for colored noises based on
the white noise representation. The integrals $J_k$, 
$2 J_{010}-J_{001} - J_{100}$, $J_{jk}-J_{kj}$, 
and $2 J_{101}-J_{110} - J_{011}$
can be conveniently calculated with discrete Fourier transformation.
While the calculations of the first two integrals are straightforward, 
the last two integrals should be handled with degenerate kernel approximation 
to reduce computational efforts.
Once the multiple integrals are calculated, the matrix exponential function 
resulted from the Magnus expansion can be evaluated with Runge-Kutta methods. 
The above procedure gives a fourth-order strong method
for single-noise SDEs and 
a third-order one for multiple-noise SDEs. 

The current approach is simple and easy to implement. Furthermore, 
although the method is derived for autonomous linear stochastic differential 
equations, it can be readily extended to general equations as well. 

\section*{Acknowledgments}

The authors acknowledge the support from the National
Natural Science Foundation of China under grant No. 21373064.

\end{document}